# Learning Clustered Representation for Complex Free Energy Landscapes

Jun Zhang[a,b], Yao-Kun Lei[a], Xing Che[c], Zhen Zhang[a,d], Yi Isaac Yang[e] and Yi Qin Gao[a,b,1]

[a]*Institute of Theoretical and Computational Chemistry, College of Chemistry and Molecular Engineering, Peking University, Beijing 100871, China.*

[b]*Biodynamic Optical Imaging Center, Peking University, Beijing 100871, China.*

[c]*Institute of Molecular Biophysics, Florida State University, Tallahassee, FL, USA.*

[d]*Department of Physics, Tangshan Normal University, Tangshan, 063000, China.*

[e]*Shenzhen Bay Laboratory, Peking University Shenzhen Graduate School, Shenzhen, 518055, China.*

[1]To whom correspondence should be addressed. Email: yangyi@szbl.ac.cn or gaoyq@pku.edu.cn

# Abstract

In this paper we first analyzed the inductive bias underlying the data scattered across complex free energy landscapes (FEL), and exploited it to train deep neural networks which yield reduced and clustered representation for the FEL. Our parametric method, called Information Distilling of Metastability (IDM), is end-to-end differentiable thus scalable to ultra-large dataset. IDM is also a clustering algorithm and is able to cluster the samples in the meantime of reducing the dimensions. Besides, as an unsupervised learning method, IDM differs from many existing dimensionality reduction and clustering methods in that it neither requires a cherry-picked distance metric nor the ground-true number of clusters, and that it can be used to unroll and zoom-in the hierarchical FEL with respect to different timescales. Through multiple experiments, we show that IDM can achieve physically meaningful representations which partition the FEL into well-defined metastable states hence are amenable for downstream tasks such as mechanism analysis and kinetic modeling.

# 1 Introduction

Along the development of rate theory in chemical physics, the most important insight of many dynamic systems is that there usually exists a separation of timescales[1-4], so that interesting events (inter-state transitions) take place on a much longer timescale (denoted as $\tau_{ts}$, the inverse of which defines the rate coefficient in physics) than the internal relaxation within the state ($\tau_{rx}$), that is, $\tau_{ts} \gg \tau_{rx}$. In other words, given the separation of timescales, each state will reach a local equilibrium within a characteristic timescale $\tau_{rx}$, but transitions to other states may occur on longer timescales (called "rare events"). This observation leads to the notion of metastability[5], and such states are termed as metastable states. A well-defined metastable state should exhibit an exponentially decayed lifetime since the escape from it is approximately a Poisson point process[3, 6]. Alternatively, from the point view of the landscape theory[7], energetically accessible configurations take up only a small fraction of phase space for molecules like proteins. Consequently, a properly defined free-energy landscape (FEL) usually consists of heavily clustered populations, on which each cluster forms a local free-energy minimum and corresponds to a metastable state[8]. Many molecular dynamic processes in chemistry and biology, e.g. chemical reaction, protein folding and ligand binding etc, can be described by such a complex free-energy landscape where long-lived metastable states are separated by kinetic bottlenecks (i.e. free-energy barriers) which impede the transitions[9].

The picture of a clustered FEL (or the metastability), which we note here as the inductive bias of FEL, is the cornerstone of many state-of-art kinetic models which deal with diffusive and complex dynamics, e.g., (discrete or coarse) master equation[10-11], transition path theory (TPT)[12-13], Markov state models (MSM)[14-16], etc. Therefore, in order to demystify the dynamic physical processes, a simplified and informative visualization of complex FEL, which is amenable for downstream tasks such as clustering, is often required by these kinetic modeling methods. Usually this is achieved by linear or non-linear dimensionality reduction techniques, e.g. principal component analysis (PCA)[17], time-lagged independent component analysis (tICA)[18-19], Isomap[20], sketch map[21] and diffusion map (DM)[22-23]. However, most of these existing methods are subjected to cherry-picked distance metrics (the choice of which is often very tricky). Furthermore, many methods only depend on the geometric representation of samples and few of them incorporate the available dynamic information, thus may not adequately capture the information of metastability. Worse still, most nonlinear dimensionality reduction methods (e.g., Isomap, DM) involve computationally prohibitive non-parametric kernels thus cannot directly scale to ultra-large dataset. Last but not the least, it is non-trivial to partition the FEL into metastable states even based on the reduced depiction obtained by these methods, and commonly the assumption of metastability can only be checked by post-mortem analysis[15].

On the other hand, learning informative representation has long been one of the most important goals for machine learning, and several recently developed deep models (e.g., variational auto-encoder[24], adversarially learned inference[25] etc) have shown great potentiality for unsupervised or semi-supervised representation learning. However, few (if any) of the state-of-art models can directly fit in the above task, since FEL inherits very different inductive biases from the data types commonly encountered in deep learning community. We now propose to unsupervisedly learn a reduced representation for FEL where different metastable states are embedded into separate clusters, without the necessity of pre-defined distance metrics or number of clusters. Our approach is based on parametric models (e.g., artificial neural networks) hence can be trained efficiently with stochastic gradient descent (SGD) on mini-batches of ultra-large dataset. More importantly, the reduced representation is jointly learned together with clustering in one shot, differing with the common practice that to cluster one needs to first reduce the dimension in a separate manner. Our method allows deep models to capture the inductive bias introduced above and extract meaningful representations directly from high-dimensional dynamic trajectories. This is achieved by maximizing the mutual information of related samples so as to distill their shared abstract content. Similar idea of learning a data representation from related observations is not new. An early work in this line can be traced back to Becker and Hinton[26], where they maximized the mutual information between the input and the average of the data representations. Co-learning has also been explored in the context of clusterings, dating back to the pioneering work of Hartigan[27]. Particularly, distilling the information between related samples has already been successfully applied for image clustering and segmentation[28]. Albeit the idea of information distilling has been most exploited in image processing, IDM, on the other hand, shows that by capturing the inductive bias of FEL, complicated physical problems can also be elegantly addressed following the same line.

# 2 Methods

## 2.1 Information Distilling of Metastability (IDM)

Our goal is to associate with each unlabeled sample $\mathbf{x}$ a label $\chi(\mathbf{x})$ indicative of its identity (i.e. the probability of $\mathbf{x}$ being in certain metastable states, or termed as "membership") without any ground truth for direction. To achieve this, consider the related problem of co-distillation[28]: given two observations $\mathbf{x}$ and $\mathbf{x}'$ belonging to the same metastable state, we seek for a function $\chi$ that captures what is in common between them. Formally, it is naturally to solve this problem by maximizing the mutual information (denoted by $I$) $I(\chi(\mathbf{x}), \chi(\mathbf{x}'))$ between their representations. However, due to the data processing inequality[29], i.e. $I(\mathbf{x}, \mathbf{x}') \geq I(\chi(\mathbf{x}), \chi(\mathbf{x}'))$, we see that this is trivially solved by choosing $\chi$ to be the identity function. To combat the problem of trivial solution, we can impose an information bottleneck over $\chi$, that is, to deliberately treat $\chi$ as a discrete classification function (or indicator function), i.e., $\chi(\mathbf{x}) \in \{1, 2, \ldots, K\}$ where $K$ is the number of clusters. As a consequence, the entropy of $\chi$ is restricted by a ceiling, and the optimized $\chi$ automatically serves as a clustering function. In practice, we consider $\chi$ belonging to the family of neural networks that terminates in a multi-logit layer, to approximate the deterministic mapping described above.

We now exploit the inductive bias of FEL introduced in the previous section, yielding the information distilling of metastability (IDM), which directly clusters the high-dimensional configurations $\mathbf{x}$ embedded on complex FEL into different metastable states. The key idea of IDM is to generate a related sample $\mathbf{x}'$ from $\mathbf{x}$ so that $\mathbf{x}'$ and $\mathbf{x}$ *almost surely* belong to the same state. According to the metastability assumption, this can be cheaply done by sampling $\mathbf{x}'$ from the temporal proximity of $\mathbf{x}$ (denoted as $\epsilon(\mathbf{x}; \tau)$) via dynamic simulation techniques such as molecular dynamics (MD)[30]. Note that $\tau$ is a hyper-parameter that defines the temporal resolution of the model, provided that motions on timescales shorter than $\tau$ will be regarded as internal relaxations. In this way, a good model should satisfy the requirement that $\chi(\mathbf{x}) = \chi(\mathbf{x}')$, which can be used as a constraint in clustering methods. However, herein we can achieve the above goal with an alternative objective function,

$$\max_\chi I_\beta(\chi(\mathbf{x}), \chi(\mathbf{x}')) \tag{1}$$

where we introduce a controlling hyper-parameter $\beta$ without loss of generality. Let $\chi_c(\mathbf{x})$ denote the $c$-th element of $\chi$ which corresponds to the probability of $\mathbf{x}$ belonging to cluster $c$, then the mutual information takes the following form,

$$I_\beta = \sum_{c,c'=1}^{K} \langle \chi_c(\mathbf{x}), \chi_{c'}(\mathbf{x'}) \rangle \ln \frac{\langle \chi_c(\mathbf{x}), \chi_{c'}(\mathbf{x'}) \rangle}{\langle \chi_c(\mathbf{x}) \rangle^\beta \langle \chi_{c'}(\mathbf{x'}) \rangle^\beta} \quad (2)$$

where $\langle \cdot \rangle$ denotes the expectation over the paired dataset $\{(\mathbf{x}, \mathbf{x'})\}$, and can be unbiasedly estimated via mini-batches of samples. Intuitively Eq. 2 means the clustering mapping of a given configuration, namely $\chi(\mathbf{x})$, should be maximally informative with its temporal neighbors. In practice, this is obtained when $\chi(\mathbf{x}) \approx \chi(\mathbf{x'})$ for arbitrary $\mathbf{x}$. Noteworthy, Eq. 2 can be conveniently maximized over mini-batches of paired samples $(\mathbf{x}, \mathbf{x'})$ with gradient-based optimization techniques.

When $\beta = 1$ Eq. 2 reduces to the Kullback-Leibler (KL) divergence between the joint distribution and the product of two marginal distributions, and we can make several remarks on the IDM objective hereby. Firstly, Eq. 1 inherits the cluster mass equalization bias[28] which actively avoids the trivial solution of categorizing all inputs into the same clusters. This can be seen by considering the degenerate case where $\beta = 1$ and the paired dataset becomes $\{(\mathbf{x}, \mathbf{x})\}$: As $\chi(\mathbf{x})$ is a deterministic function of $\mathbf{x}$, $I_1(\chi(\mathbf{x}), \chi(\mathbf{x})) = H(\chi(\mathbf{x})) - H(\chi(\mathbf{x})|\chi(\mathbf{x}))$. Since the conditional entropy $H(\chi(\mathbf{x})|\chi(\mathbf{x})) = 0$, Eq. 1 reduces to $\max_\chi H(\chi(\mathbf{x}))$, where $H(\chi(\mathbf{x}))$ denotes entropy over the entire dataset $\{\mathbf{x}\}$,

$$H(\chi(\mathbf{x})) = -\sum_{c=1}^{K} \langle \chi_c(\mathbf{x}) \rangle_{\{\mathbf{x}\}} \ln \langle \chi_c(\mathbf{x}) \rangle_{\{\mathbf{x}\}} \quad (3)$$

Consequently, the optimal $\chi$ tends to partition the mass of the data equally among all the clusters. Secondly, we can explicitly control the balance of clusters produced by $\chi$ by tuning the controlling hyper-parameter $\beta$. Note that in the form of KL-divergence,

$$I_\beta(\chi(\mathbf{x}), \chi(\mathbf{x'})) = I_1(\chi(\mathbf{x}), \chi(\mathbf{x'})) + (\beta - 1)\left(H(\chi(\mathbf{x})) + H(\chi(\mathbf{x'}))\right) \quad (4)$$

so setting $\beta > 1$ forces $\chi$ to produce more evenly spread clusters while $\beta < 1$ allows more imbalanced ones. High $\beta$ encourages the network to exploit its full expressive capacity, thus overcoming poor initialization or heavily biased initial cluster assignments. In all the experiments done in this work, we employ a decaying schedule to relax $\beta$ from a higher value to unity during training (see SI for more details). This schedule encourages $\chi$ to form as many nascent clusters as possible in the early stage and to coalesce degenerate clusters in the later stage. In fact, a periodical schedule of $\beta$ will work as simulated annealing and may perform even better in more complicated cases, and we leave this idea for future research.

Furthermore, in an entirely unsupervised setting, the number of ground-truth clusters is unknown. IDM is naturally devoid of this problem as the conditional entropy in the mutual information tends to shrink unnecessary modes in the clustering, as long as $\beta$ is not too high. This coercing effect will be more pronounced in combination with the regularization technique described later (Section 2.3). Therefore, IDM only requires the user to specify an upper limit on the number of output clusters. In practice the performance of IDM can be improved by setting the number of output clusters greater than the ground truth as the network becomes more expressive[28].

**2.2 IDM via deep learning**

As introduced above, a simplified visualization of complex free energy landscape is often needed in order to understand the dynamic physical processes. We show here that by virtue of IDM, the reduced representation can be jointly learned together with clustering. Specifically, our objective is to find a function $G_\theta(\mathbf{x}): \mathbb{R}^D \to \mathbb{R}^d$ ($D > d$), where $\mathbb{R}^d$ is defined as the *matching space* and serves as the reduced representation of the FEL. This goal is straightforwardly achieved by deliberately employing a special deep neural network architecture, Matching Network[31-32]. The Matching Network consists of two twin networks (for uncluttered notation, we denote the parameters of both networks with a same symbol $\theta$): one is $G_\theta(\mathbf{x})$ as defined above; the other is $g_\theta(\mathbf{c})$ to map a certain code vector or a (pseudo-)prototype, $\mathbf{c}$, to the same matching space. In other words, we map the data and the prototypes both to the matching space $\mathbb{R}^d$ with $G_\theta$ and $g_\theta$, respectively. In practice, if without any *a prior* knowledge, we can use one-hot vectors (Fig. 1, and see SI Text 1.10 for more details) as the prototypes. If available structures representing different metastable states are known, they can be directly used as prototypes, and $g_\theta$ in this case is degenerate to $G_\theta$.

On the matching space, the symmetric similarity score, $S(\mathbf{x}, \mathbf{c}) = S(\mathbf{c}, \mathbf{x})$, between the projected sample $G_\theta(\mathbf{x})$ and each projected prototype (serving as cluster centroid) $g_\theta(\mathbf{c})$ can be simply defined, e.g., via distance-based metrics[33] or the dot product as in the attention mechanism[34] (see more details of similarity scores in SI). For example, we can adopt the attention mechanism to calculate $S(\mathbf{x}, \mathbf{c})$ (Eq. S9), then use it as the multi-logit to output the probability of $\mathbf{x}$ being to cluster $j$ centered at $g_\theta(\mathbf{c}_j)$, with the temperature factor $\gamma$ tuned as a hyper-parameter,

$$P(\mathbf{x} \in \text{cluster } j) = \chi_j(\mathbf{x}) = \frac{\exp \gamma S(\mathbf{x}, \mathbf{c}_j)}{\sum_{i=1}^K \exp \gamma S(\mathbf{x}, \mathbf{c}_i)} \tag{5}$$

Therefore, training of $G_\theta$ and $g_\theta$ can be directly done by optimizing the objective function Eq. 2 with $\chi(\mathbf{x})$ substituted by Eq. 5. For applications where a discrete label for samples is preferred, we can simply achieve this by "hardening" the soft clustering results, that is, labeling $\mathbf{x}$ with $j$ that maximizes $\chi_j(\mathbf{x})$.

**2.3 Regularization and training**

In order to stabilize the training process, we regularized our models with two additional techniques (see SI Text 1.1 and 1.2 for more details): 1) adaptive gradient clipping to avoid exploding gradient, and 2) performing auxiliary tasks inspired by multi-task learning[35]. In our work the auxiliary task is naturally set to reconstruct the high-dimensional input from the reduced matching space (Fig. 1). Next, we describe how to prepare the training data for mini-batch optimization. The strategy we take is similar to the task sampling in meta-learning[36-37], curriculum learning[38] and few-shot learning[31]. Given a dataset, $\{\mathbf{x}\}_F$, containing samples scattered across the FEL and being randomly shuffled, for each training step, we pop the first $s$ samples from the queued $\{\mathbf{x}\}_F$ as a seeding set $\{\mathbf{x}\}_S$, and perform farthest point sampling[39] (using crude distance metric, e.g. RMSD) within the entire $\{\mathbf{x}\}_F$ to expand the seeding set up to $\{\mathbf{x}\}_N$ with size $N > s$. Next for each $\mathbf{x} \in \{\mathbf{x}\}_N$, we perform neighbor sampling by MD simulations initialized at $\mathbf{x}$ and collect configurations within time $\tau$ as $\epsilon(\mathbf{x}; \tau)$. Finally, we feed $\{\mathbf{x}\}_N$ and the paired $\{\epsilon(\mathbf{x}; \tau)\}_N$ into the model and perform one step of SGD with Adam[40] (see SI for more training settings). One epoch of training is done when no more samples are left queued in $\{\mathbf{x}\}_F$. The convergence is achieved when the training objective (Eq. 2 or Eq. S3) becomes steady and no longer changes significantly.

# 3 Results

## 3.1 Dimensionality reduction

We first illustrated the performance of IDM on a numerical model potential[21] (see SI for the model setups and training details), which shares many features common to real-world free-energy landscapes (Fig. 2A, panel 1). To fully specify the configurations while taking into account of the periodicity, we assign each sample a 6-dimensional vector $\mathbf{x} = \cup_{i=x,y,z}\{\cos i, \sin i\}$ and use it as the input to the dimensionality reduction algorithms. Figure 2A (panel 2) shows that IDM ($\tau = 10$) clearly projects all the eight local energy minima onto the matching space and yields a clustered and well-aligned embedding. We also tested several other manifold learning algorithms including PCA, tICA and PCA for comparison (see SI Text for more details). Note that this model potential is periodic in the $(x, y, z)$ dimensions, hence it cannot be mapped isometrically to a linear two dimensional space[21]. Although PCA and DM (Fig. 2A) also yield a clustered projection of the potential energy surface (while tICA fails), the organization of the resultant clusters becomes obscure to interpret. In contrast, IDM is able to preserve most transition pathways by breaking only a few connections between basins. Indeed, IDM learns to unroll the periodical box rather than simply squashing the box onto the plane, and the resulting embedding clearly sketches the original structure or shape of the configuration space. Such feature renders IDM appealing potentiality of guiding enhanced sampling methods like umbrella sampling[41] and metadynamics[42].

Next, we tried IDM on alanine dipeptide (Ala2; see SI for data sources and training details) to see whether it can learn a reduced but meaningful representation from raw coordinates of biomolecules. In order to retain the trans-rotational invariance, we chose all the 45 pairwise (properly normalized) distances between heavy atoms as the input vector $\mathbf{x}$. Figure 2B (panel 2) shows that IDM ($\tau = 20$ ps) clearly projects the high-dimensional vector $\mathbf{x}$ onto 4 distinct free-energy minima, in agreement with our knowledge that Ala2 exhibits 4 metastable conformers with respect to the two torsional angles $(\phi, \varphi)$ (Fig. 2B, panel 1). Furthermore, IDM again maximally preserves the transition paths connecting metastable states (Fig. 2B). Comparable result is only obtained by tICA out of other methods we tested. By virtue of the expressive power of deep neural networks, these two examples demonstrate the ability of IDM to extract useful representations from crude coordinates of the system without carefully handcrafted order parameters.

## 3.2 Clustering

As introduced, IDM distinguish itself from other dimensionality reduction methods in that it simultaneously clusters the data during training. More importantly, one does not need to specify the exact number of clusters in IDM. Instead, we only need to provide an estimation for the upper bound, denoted by $K$. For ease of comparison, we only reported the hard clustering results if not stated otherwise. We found that IDM performs robustly as long as $K$ is large enough (which algorithmically means that the neural network has adequate capacity). For instance, we chose $K = 16$ for both cases studied above. In the numerical model system, only 8 out of 16 clusters were substantially populated after training of IDM is done, agreeing with the fact that there are 8 metastable states. For Ala2, only 4 out of the 16 prescribed clusters were essentially occupied, also well according with the ground truth. This result shows the effectiveness of IDM compared to many other clustering methods (e.g. KMeans) in that IDM is a completely unsupervised algorithm subjected to least manual interference.

To interpret the clustering results, we visualized the samples drawn from the numerical model potential (Fig. 2A) and Ala2 (Fig. 2B) according to their cluster identity on a meaningful representation. Figure 2A shows that the eight clusters for the model potential obtained by IDM correspond exactly to the potential energy minima. Figure 2B concludes that IDM clusters configurations of Ala2 into 4 free energy minima corresponding to the 4 different *cis-/trans*-isomers of the $(\phi, \varphi)$ torsions.

Furthermore, since we have access to the reference labels for these two well-benchmarked systems, we can quantitatively assess the performance of different clustering approaches. Two different metrics, clustering accuracy (ACC) and normalized mutual information (NMI) (see definitions in SI), were adopted (Table 1). In addition to direct clustering by IDM, we performed KMeans (setting $K$ to be the ground-true values) on the reduced embedding achieved by IDM, PCA, tICA and DM, respectively. Table 1 shows that the direct IDM clustering consistently outperforms other combinatory strategies. Besides, it can be noticed that the performance of KMeans is slightly improved based on the IDM representation compared to other reduced embeddings. These results demonstrate that IDM is able to project the complex FEL onto a clustered representation which preserves important physical properties of the system, meanwhile partitioning the FEL into well-defined metastable states.

**3.3 Kinetic modeling**

Many kinetic modeling approaches are derived from the Markovian master equation (e.g. TPT and MSM), which requires proper clustering of configurations so that transitions between different cells are approximately Markovian, that is, the escape from each cell is Poisson point process[6, 11]. To physicists, a natural way to ensure Markovian transitions is to correspond the cells to the metastable states, because the separation of timescales guarantees the escape event a Poisson process. We thus expect these multi-state kinetic modeling approaches can benefit from IDM which by design intends to cluster configurations into metastable states. But before drawing the conclusion, one needs to first check the Markovian assumption. Figure S1A shows the lifetime distribution of the four clusters obtained by IDM for Ala2: every cluster exhibits an exponentially decayed lifetime indicating a Poisson escaping behavior (in other words, each cluster is indeed a metastable state). Similar results were obtained for the numerical model system (Fig. S2A), which allows the analysis of the long-timescale behavior of the dynamic systems. In contrast, when different metastable states are mixed (e.g., as Fig. S1B, yielded by tICA+KMeans[18-19]), the lifetime distribution of the resulting clusters could violate the desired exponential decay.

As an illustrative example, we estimated a coarse master equation[10] for Ala2 based on IDM, whose propagating equation reads,

$$\chi(\mathbf{x}_{t+n\tau}) = \exp(n\mathbf{K}_\tau) \chi(\mathbf{x}_t) \qquad (6)$$

where $\exp(n\mathbf{K}_\tau)$ is a matrix exponential, and $\mathbf{K}_\tau$ is a dimensionless transition rate matrix corresponding to a Markovian master equation, the time resolution $\tau$ of which is coarse-grained to be the same as IDM ($\tau = 10$ ps). Equation 6 holds when $n \gg 1$ under metastability assumption. The transition rate matrix $\mathbf{K}_\tau$ is estimated by maximizing the log-likelihood objective $\mathcal{L}$ in Eq. 7 constrained by mass conservation and detailed balance[10] via an efficient evolution-based optimization technique[43] (see SI for more details).

$$\mathcal{L} = \sum_t \chi(\mathbf{x}_{t+n\tau})^{\mathrm{T}} \cdot \ln[\exp(n\mathbf{K}_\tau) \chi(\mathbf{x}_t)] \qquad (7)$$

where the logarithm is performed element-wisely. Note that since Eq. 7 is directly fed with the soft clustering labels $\chi(\mathbf{x})$, it is equivalent to the negative cross-entropy which is a cost-sensitive objective for training[44]: Unlike conventional MSM which usually fully discretizes (or called hard clusters) the trajectory, the likelihood function here emphasizes more on an accurate prediction of highly-identifiable configurations (those near the bottom of metastable states) rather than the ones with higher uncertainties (those located on the boundaries between metastable states). Such a cost-sensitive objective greatly reduces the side-effect of hard-clustering such as re-crossing biases encountered in MSM[45]. Figure 4 shows the extracted transition pathways and the associated rates, revealing a sharper transition of $\varphi$ torsion (between states 1 and 3, or states 2 and 4) than the $\phi$ torsion (between states 1 and 2, or states 3 and 4). Additionally, the coarse master equation for the numerical model was obtained in a similar way (see SI for more details), according to which we drew the eigenspectrum of the rate matrix (Fig. S2B) and extracted the main transition pathways (Fig. S2C): As expected, the resulting rate matrix only contains statistically non-zero rates between physically adjacent states.

We also noticed that clusters obtained by IDM are generally more stable and exhibit longer lifetime compared to the commonly adopted combinatory projection-clustering method (Fig. S1). This effect results from the fact that IDM defines boundaries between clusters according to separation of timescales rather than a geometric cutoff as adopted in KMeans. Consequently, IDM clusters are less vulnerable to the notorious "fast re-crossing" issues pronounced in the state-of-art method[16] which causes severe underestimate of the lifetime (or overestimate of transition rates). Altogether, the above analyses show that IDM can yield clusters and representations that are more amenable for downstream tasks like kinetic modeling.

**3.4 Zooming-in hierarchical free-energy landscape**

Many complex physical and biological processes can be described by hierarchical FEL. Specifically, the identity of metastable states and the slow inter-state transition processes depend on the timescales at which one inspects the system. Put it in another way, the representations of FEL may vary at different time resolutions. Trained upon $N$ decreasing resolutions ($\tau_1 \gg \tau_2 \gg \cdots \gg \tau_N$), IDM allows us to zoom-in the FEL with increasing time resolutions. By doing so, we are actually performing a top-down divisive clustering which remains challenging for any other algorithms, hence we term this approach as divisive IDM (see more details about divisive IDM in SI). Noteworthy, divisive IDM is able to extract the hierarchy of the FEL according to timescales rather than geometry-based metrics.

We performed divisive IDM on Ala2 for illustration (Fig. 5). Three timescales were chosen for training: (1) $\tau = 200$ ps, (2) $\tau = 20$ ps, and (3) $\tau = 2$ ps. On the longest timescale ($\tau = 200$ ps), IDM partitions all conformations into two metastable states (Fig. 5A, panel 1), corresponding to the *cis/trans*-isomers of the torsional angle $\phi$. This is in good agreement with previous kinetic modeling result that isomerization of $\phi$ is much slower than $\varphi$ (Fig. 4). As expected, the reduced representation obtained by IDM at this timescale only preserves two distinguishable metastable states (Fig. 5B, panel 1). When progressing to a smaller timescale of 20 ps, IDM further divides the conformations into four clusters, as reported in previous sections (panel 2 in Fig. 5A), and the reconstructed FEL now consists of 4 distinguishable metastable states (panel 2 in Fig. 5B). Moreover, divisive IDM allows us to track the lineage of the hierarchy, so we can plot the dendrogram of the clusters or metastable states (Fig. 5C). Clusters at each level are indexed with a prefix indicating the level of resolution. From Fig. 5 we can see that by increasing the resolution from $\tau = 200$ ps to 20 ps, isomerization of $\varphi$ is identified by IDM also as slow inter-state transition processes. Following the same line, when we continue to tune $\tau$ down to 2 ps, Cluster 2-1 is further divided into 2 metastable states (3-1 and 3-2, Fig. 5C). At this very small timescale, IDM categorizes all the conformers of Ala2 into 5 metastable states (Fig. 5A, panel 3) and learns a reduced representation accordingly (Fig. 5B, panel 3). These 5 states indeed correspond to the well-known metastable conformations of Ala2, including $\beta$ (3-1), PP$_{II}$ (3-2), $\alpha_R$ (3-3) and $\alpha_L$ (3-5). In summary, this example shows that divisive IDM allows us to zoom-in the FEL with increasing time resolutions and track the hierarchy of metastable states accordingly.

### 3.5 Unrolling folding landscape of proteins

Finally, we present an application of IDM on a fast-folding protein TrpCage, trying to reveal more molecular details of the folding mechanisms (see SI for the training details). Since the configurations of a protein are largely determined by the backbone torsional angles, we first collect all backbone torsions, $\{\Phi_i, \Psi_i\}_i$ where $i$ runs over every residue (20 residues in total), then transform them into a 76-dimensional torus vector,

$$\mathbf{x} = \cup_i \{\cos \Phi_i, \sin \Phi_i, \cos \Psi_i, \sin \Psi_i\}$$

which serves as the input vector to IDM. We chose $K = 32$ during training and performed divisive IDM at three time resolutions: (1) $\tau = 400$ ns, (2) $\tau = 40$ ns and (3) $\tau = 4$ ns, respectively. Models were trained upon the MD trajectories contributed by K. Lindorff-Larsen et al[46], which contain over 500,000 samples.

On the longest timescale we chose (i.e., $\tau = 400$ ns), IDM projects the FEL onto a reduced representation consisting of 2 distinguishable metastable states (Fig. 6A, panel 1) and categorizes the protein conformers into 2 clusters accordingly. In order to interpret the results, for each cluster we calculated the averaged root-mean-squared-deviation (RMSD) w.r.t. the native structure and the fraction of native contacts ($Q$-value), respectively (Table S1). Cluster 1-1, with high $Q$-value and low RMSD, is identified as the native state; whereas Cluster 1-2 corresponds to the denatured state. In line with intuitions, the native state forms a narrow and sharp local minimum on the IDM embedding surface, while the unfolded states spread more widely over the space (Fig. 6A, panel 1). This two-state picture agrees well with the observations that the folding/unfolding events of proteins can be viewed and measured as a two-state kinetic process on relatively long timescales.

An intriguing finding is that, even if we increase the time resolution up to $\tau = 4$ ns, we still do not observe any further division of the native state (Figs. 6B and 6C), indicating no discernible slow relaxations within the native state. This result strongly corroborates the common belief that proteins exhibit a stable and well-defined native structure whose fluctuations (and the conformational entropy) are rather limited. In contrast, some slow relaxation processes within the denatured state can be reconsidered as inter-state transitions. Consequently, more unfolded metastable states can be identified. Specifically, at the resolution of $\tau = 40$ ns, the unfolded state (Cluster 1-2) divides into two distinguishable clusters (indexed by 2-2 and 2-3, respectively; Fig. 6B). From the embedding surface yielded by IDM (Fig. 6B, panel 2) we find that State 2-2 is more closely connected to the native state while State 2-3 is farther away, implying that State 2-2 is likely to be located on the folding pathway (which may contain the folding bottleneck) while State 2-3 may be less relevant to folding. This hypothesis is confirmed by the IDM embedding achieved at $\tau = 4$ ns (Fig. 6A, panel 3): Cluster 2-2 bifurcates into States 3-2 and 3-3 (Fig. 6B), and the former corresponds to a folding bottleneck where some characteristic contacts start to form (e.g., the $\alpha$–helix), while the latter is composed of extended random coils which represent the fully denatured states (Fig. 6C). On the other hand, Cluster 2-3 are divided into 4 metastable states (indexed from 3-4 to 3-7; Fig. 6B): Most of them exhibit some locally formed contacts and structural patterns absent in the native structure (Fig. 6C), but these potentially folding traps only constitute a small fraction of the denatured conformation ensemble, as expected for a fast-folding protein. Additionally, we find it interesting that, as the time resolution increases and the metastable states assigned by IDM become devoid of slower intra-state relaxations, the boundaries between clusters become sharper on the reduced embedding surface (Fig. 6A, from panel 1 to panel 3). Accordingly, on a low-resolution representation of FEL (as in Fig. 6A, panel 1), the inter-state kinetics can be described by a relatively low free-energy barrier and a slow diffusion term in the pre-exponential factor. In contrast, on a high-resolution representation (as in Fig. 6A, panel 3), the inter-state kinetics are characterized by a higher free-energy barrier but a faster diffusion term. These results echo the well-known funnel landscape theory[47], and demonstrate that we can exploit IDM to unroll and project the funnel energy landscape of protein in a hierarchical manner. Therefore, IDM may be applied to shed more light on the mechanisms of protein folding.

# 4 Concluding remarks

Supercomputing gives access to large amounts of high-dimensional simulation trajectories of complex physical and biological processes of interest. However, in order to extract relevant information and reveal the key factors that determine the underlying mechanisms, a reduced description of the high-dimensional data is often desired. In other words, a highly expressive and flexible method is needed to learn a meaningful reduced representation for the FEL. In this regard, deep learning seems promising to offer a possible solution. From the perspective of deep learning, capturing the inductive biases within the data is the foundation of successful deep models. Examples include the success of convolutional neural networks[48] over images and recurrent neural networks[49] over texts etc. Therefore, if one wants to characterize the complex FEL with a deep model, the model should by itself exploit the inductive bias of FEL, which has not been adequately taken account of.

In this paper, we took advantage of the inductive bias that a correct and useful representation of FEL should be composed of clustered population due to metastability, and developed a representation learning method, IDM. By distilling the information between temporally close samples via deep neural networks, IDM naturally preserves the metastability of samples and yields clustered representation of FEL. Despite the fact that distorting the topology of the original space is inevitable for dimensionality reduction, IDM may manage to partially keep the correct kinetic connectivity between different metastable states because similarity is defined according to temporal proximity. Moreover, IDM has several important algorithmic merits: Foremost, IDM is based on flexible parametric models (neural networks) hence is readily scalable to ultra-large dataset; secondly, IDM does not require a pre-defined distance metric or similarity kernel but rather learns it automatically.

IDM is also a self-contained clustering algorithm that yields a soft partitioning of the FEL without need of further processing. It is shown that the clusters obtained by IDM indeed correspond to metastable states whose lifetime exponentially decays and the escaping events are Poisson point processes. Remarkably, IDM does not require a reference to the exact number of ground-true clusters. Instead, IDM robustly clusters the data within a maximal allowed number of clusters specified by users. Besides, IDM yields soft clusters, which are preferred in many scenarios to hard ones. These attributes allow us to build reliable kinetic models such as coarse master equation and TPT to quantitatively investigate the dynamic processes of interest. Moreover, once is trained at different time resolutions, IDM is able to unroll and zoom-in the hierarchical FEL for complex dynamic systems like protein folding. In this sense, IDM is a novel algorithm that can be used to perform divisive clustering and dimensionality reduction with respect to varying timescales. As an example, we employed IDM to analyze the MD simulation trajectories of TrpCage, and revealed rich molecular details of the folding dynamics and unrolled the hierarchy of the folding landscape, showing that even for mini-proteins there can be diverse folding intermediates and intricate folding pathways.

In addition to applications presented in this paper, we expect IDM to be useful in many other tasks since it is end-to-end differentiable and suitable for online learning settings. For example, guiding MD simulations with IDM embedding as collective variables can be an interesting direction for future research. We thus expect IDM along with the theory and optimization techniques behind it to find wide applications in theoretical studies of complex physical, chemical and biological processes.

**Table 1.** Comparison of clustering performance

| Methods | Numerical | | Ala2 | |
|---|---|---|---|---|
| | ACC[a] | NMI[b] | ACC | NMI |
| IDM | **0.999** | **0.995** | **>0.99** | **0.95** |
| IDM+KMeans | 0.998 | 0.993 | 0.83 | 0.65 |
| PCA+KMeans | 0.998 | 0.993 | 0.55 | 0.44 |
| tICA+KMeans | 0.992 | 0.98 | 0.73 | 0.53 |
| DM+KMeans | 0.997 | 0.992 | 0.81 | 0.62 |

a.  ACC is the clustering accuracy. The best performance is shown in bold.

b.  NMI is the normalized mutual information. The best performance is shown in bold.

**Figure Legends and Captions**

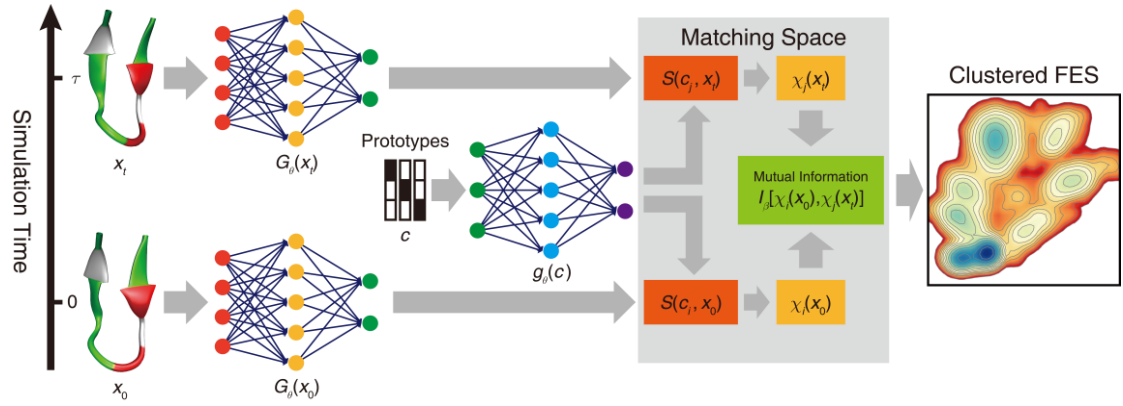

**Figure 1.** Illustrative working flow of IDM. High-dimensional input vectors (**x**) and prototypes (**c**) are projected into images in the matching space by $G_\theta$ (colored squares) and $g_\theta$ (colored circles), respectively. Similarity score between **x** and **c**, $S(\mathbf{x}, \mathbf{c})$, leads to clustering output $\chi(\mathbf{x})$ which is further used for Mutual Information ($I_\beta$) maximization.

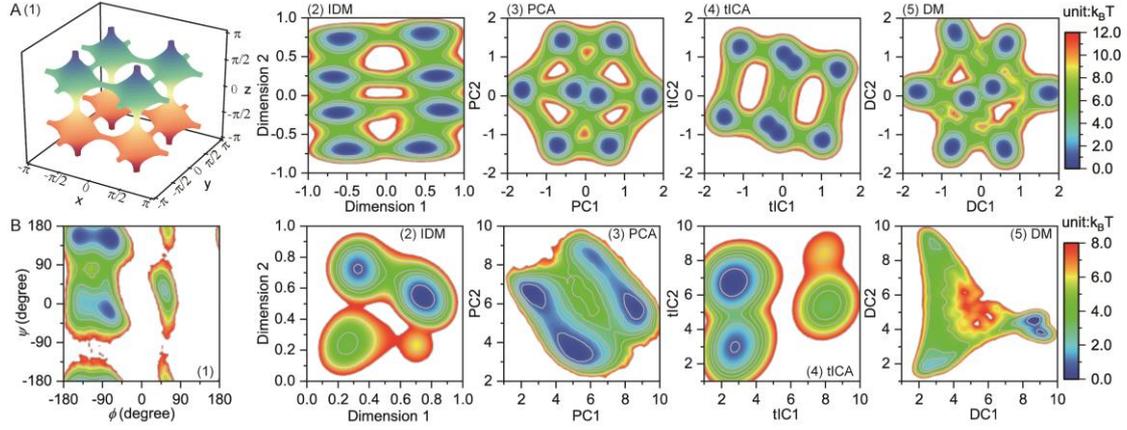

**Figure 2.** Reduced representations of free-energy landscapes for: (A) a numerical periodic potential ( $V(x,y,z) = \exp[3(3 - \sin^4 x - \sin^4 y - \sin^4 z)] - 1$ ; an iso-value surface corresponding to $\exp(-V/k_BT) = 0.01$ at temperature $k_BT$=0.25($e^3 - 1$) is shown in A1), and (B) Ala-dipeptide (a common projected visualization w.r.t. backbone torsions is shown in B1). Results yielded by different methods are shown in different column panels respectively: IDM (2), PCA (3), tICA (4) and diffusion map (5). Embedding dimensions for IDM and DM are linearly scaled for concise display.

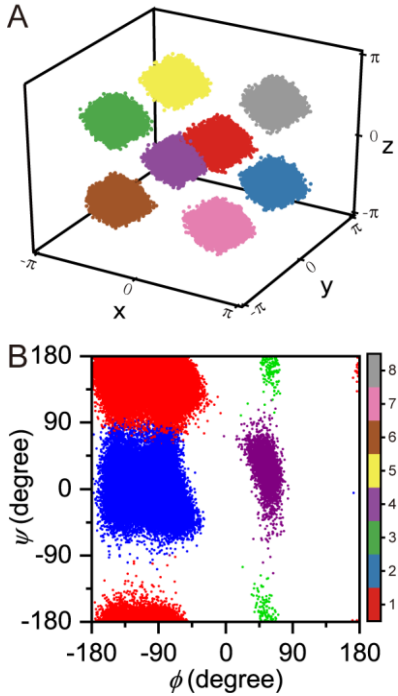

**Figure 3.** Clustering results obtained by IDM for the numerical potential (A) and Ala2 (B). Different metastable states are indexed and colored according to the color bar.

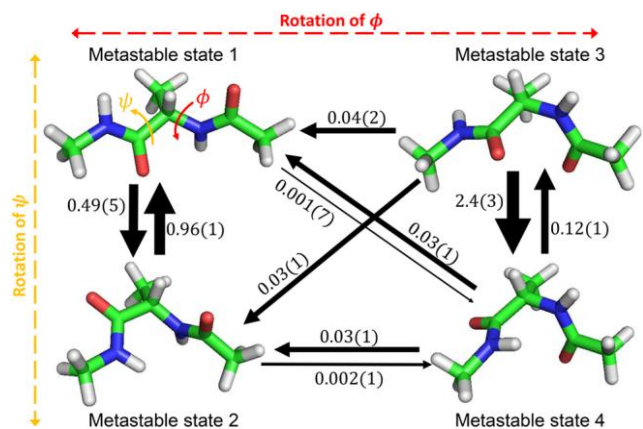

**Figure 4.** Kinetic modelling of Ala2 based on IDM (time resolution $\tau = 10$ ps). Arrows denote transition pathways, the numbers on top of the solid arrows are the dimensionless transition rates ($K_\tau$). Only transition pathways with statistically non-zero rates are shown. Indices of metastable states are the same as in Figure 3. Dashed arrows indicate the isomerization of torsional angles ($\varphi$ or $\phi$) which are marked out in the upper-left panel.

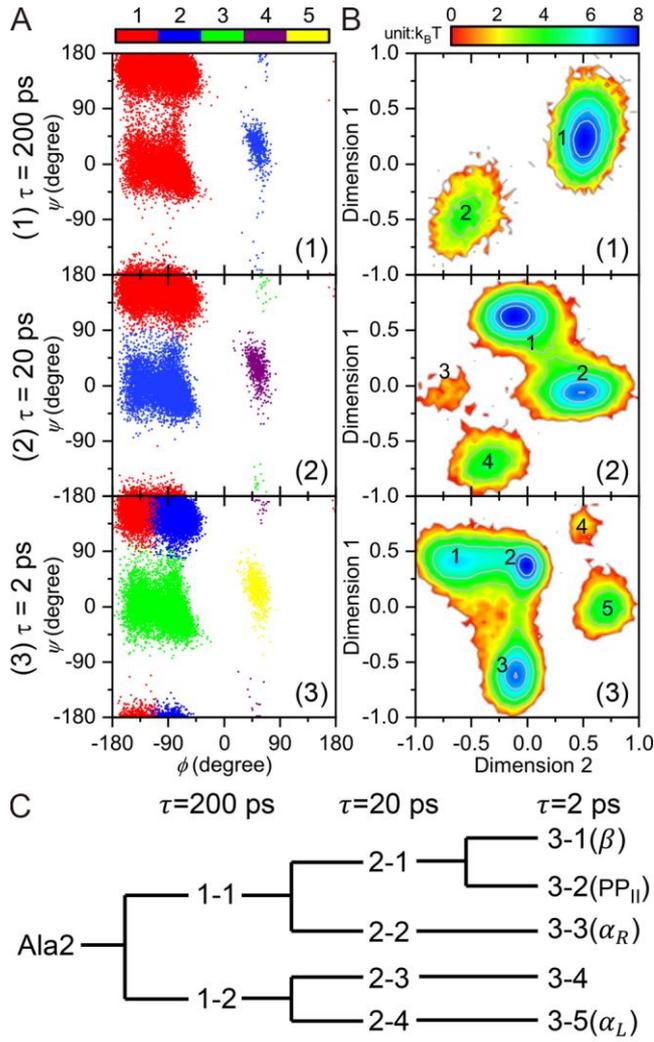

**Figure 5.** Divisive IDM performed on Ala2. (A) Clustering results obtained by IDM on different timescales shown on projected dimensions of $(\phi, \varphi)$. Clusters are colored and indexed according to the color bar. (B) Reduced representations of FEL yielded by IDM on different timescales. Metastable states are indexed in accordance with panel A. Different panels correspond to different timescales: (1) $\tau = 200$ ps, (2) $\tau = 20$ ps, and (3) $\tau = 2$ ps. (C) The dendrogram tracking the hierarchy of different metastable states identified by IDM.

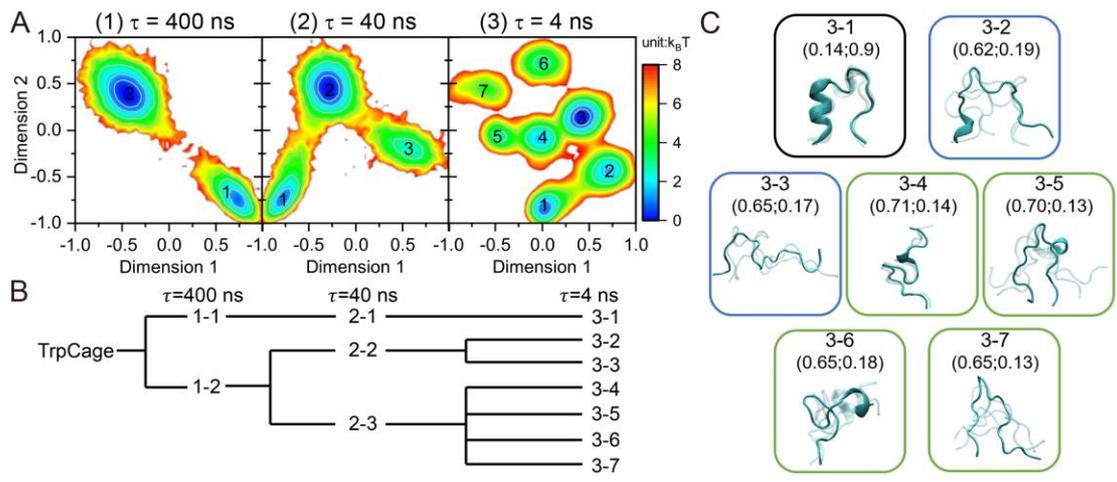

**Figure 6.** Unrolling the folding landscape of TrpCage. (A) Reduced free-energy landscape of TrpCage chignolin obtained by IDM on different timescales. From left to right: (1) $\tau = 400$ ns, (2) $\tau = 40$ ns, and (3) $\tau = 4$ ns. Metastable states are indexed accordingly. (B) The dendrogram tracking the hierarchy of different metastable states identified by IDM. (C) Representative structures of different metastable states, including folded state (3-1 in black box), folding intermediates (3-2 and 3-3 in blue boxes), and denatured structures with non-native local patterns (3-4 to 3-7 in green boxes). For each metastable state, 3 structures to the nearest of the centroid are superimposed. Numbers in parentheses are RMSD (unit: nm) and $Q$-value respectively.

## Associated Content

**Supplementary Information**

Detailed methods, system setups, model architecture, training details and additional results with associated figures are contained in the Supplementary Information.

## Acknowledgments

The authors thank Wen-Jun Xie for useful discussions. The authors also thank Prof. D.E. Shaw for sharing their MD trajectories. This research is financially supported by This research is financially supported by National Natural Science Foundation of China (21573006, 21821004, 21873007) and National Key R&D Program of China (2017YFA0204702).

## Author Information

**Corresponding authors:**

Yi Qin Gao, College of Chemistry and Molecular Engineering, Peking University, Beijing, 100871, China. Phone: 86-10-6275-2431. Email: gaoyq@pku.edu.cn

Yi Isaac Yang, Shenzhen Bay Laboratory, Peking University Shenzhen Graduate School, Shenzhen, 518055, China. Email: yangyi@szbl.ac.cn

**Author contributions:** J.Z., Y-K.L, X.C. Z.Z., Y.I.Y. and Y.Q.G. designed research; J.Z., X.C. and Y-K.L performed research; J.Z., Y-K.L, X.C. Z.Z. and Y.I.Y. analyzed the data; J.Z., Y-K.L, X.C. Z.Z., Y.I.Y. and Y.Q.G. wrote the paper.

**Notes:** The authors declare no conflict of interest.

## References:

1. Hänggi, P.; Talkner, P.; Borkovec, M., Reaction-rate theory: fifty years after Kramers. *Reviews of modern physics* **1990,** *62* (2), 251.
2. Klippenstein, S. J.; Pande, V. S.; Truhlar, D. G., Chemical kinetics and mechanisms of complex systems: A perspective on recent theoretical advances. *Journal of the American Chemical Society* **2014,** *136* (2), 528-546.
3. Peters, B., *Reaction rate theory and rare events*. Elsevier: 2017.
4. Kramers, H. A., Brownian motion in a field of force and the diffusion model of chemical reactions. *Physica* **1940,** *7* (4), 284-304.
5. Olivieri, E.; Vares, M. E., *Large deviations and metastability*. Cambridge

University Press: 2005; Vol. 100.

6. Zwanzig, R., From classical dynamics to continuous time random walks. *Journal of Statistical Physics* **1983,** *30* (2), 255-262.

7. Wales, D., *Energy landscapes: Applications to clusters, biomolecules and glasses*. Cambridge University Press: 2003.

8. Hegger, R.; Altis, A.; Nguyen, P. H.; Stock, G., How complex is the dynamics of peptide folding? *Physical review letters* **2007,** *98* (2), 028102.

9. Amadei, A.; Linssen, A. B.; Berendsen, H. J., Essential dynamics of proteins. *Proteins: Structure, Function, and Bioinformatics* **1993,** *17* (4), 412-425.

10. Hummer, G., Position-dependent diffusion coefficients and free energies from Bayesian analysis of equilibrium and replica molecular dynamics simulations. *New Journal of Physics* **2005,** *7* (1), 34.

11. Buchete, N.-V.; Hummer, G., Coarse master equations for peptide folding dynamics. *The Journal of Physical Chemistry B* **2008,** *112* (19), 6057-6069.

12. Metzner, P.; Schütte, C.; Vanden-Eijnden, E., Transition path theory for Markov jump processes. *Multiscale Modeling & Simulation* **2009,** *7* (3), 1192-1219.

13. Vanden-Eijnden, E., Transition-path theory and path-finding algorithms for the study of rare events. *Annual review of physical chemistry* **2010,** *61*, 391-420.

14. Chodera, J. D.; Swope, W. C.; Pitera, J. W.; Dill, K. A., Long-time protein folding dynamics from short-time molecular dynamics simulations. *Multiscale Modeling & Simulation* **2006,** *5* (4), 1214-1226.

15. Chodera, J. D.; Singhal, N.; Pande, V. S.; Dill, K. A.; Swope, W. C., Automatic discovery of metastable states for the construction of Markov models of macromolecular conformational dynamics. *The Journal of chemical physics* **2007,** *126* (15), 04B616.

16. Chodera, J. D.; Noé, F., Markov state models of biomolecular conformational dynamics. *Current opinion in structural biology* **2014,** *25*, 135-144.

17. Jolliffe, I., *Principal component analysis*. Springer: 2011.

18. Schwantes, C. R.; Pande, V. S., Improvements in Markov state model construction reveal many non-native interactions in the folding of NTL9. *Journal of chemical theory and computation* **2013,** *9* (4), 2000-2009.

19. Pérez-Hernández, G.; Paul, F.; Giorgino, T.; De Fabritiis, G.; Noé, F., Identification of slow molecular order parameters for Markov model construction. *The Journal*

*of chemical physics* **2013,** *139* (1), 07B604_1.

20. Balasubramanian, M.; Schwartz, E. L., The isomap algorithm and topological stability. *Science* **2002,** *295* (5552), 7-7.

21. Ceriotti, M.; Tribello, G. A.; Parrinello, M., Simplifying the representation of complex free-energy landscapes using sketch-map. *Proceedings of the National Academy of Sciences* **2011,** *108* (32), 13023-13028.

22. Nadler, B.; Lafon, S.; Kevrekidis, I.; Coifman, R. R. In *Diffusion maps, spectral clustering and eigenfunctions of Fokker-Planck operators*, Advances in neural information processing systems, 2006; pp 955-962.

23. Ferguson, A. L.; Panagiotopoulos, A. Z.; Kevrekidis, I. G.; Debenedetti, P. G., Nonlinear dimensionality reduction in molecular simulation: The diffusion map approach. *Chemical Physics Letters* **2011,** *509* (1-3), 1-11.

24. Kingma, D. P.; Welling, M., Auto-encoding variational bayes. *arXiv preprint arXiv:1312.6114* **2013**.

25. Dumoulin, V.; Belghazi, I.; Poole, B.; Mastropietro, O.; Lamb, A.; Arjovsky, M.; Courville, A., Adversarially learned inference. *arXiv preprint arXiv:1606.00704* **2016**.

26. Becker, S.; Hinton, G. E., Self-organizing neural network that discovers surfaces in random-dot stereograms. *Nature* **1992,** *355* (6356), 161.

27. Hartigan, J. A., Direct clustering of a data matrix. *Journal of the american statistical association* **1972,** *67* (337), 123-129.

28. Ji, X.; Henriques, J. F.; Vedaldi, A., Invariant information distillation for unsupervised image segmentation and clustering. *arXiv preprint arXiv:1807.06653* **2018**.

29. Cover, T. M.; Thomas, J. A., *Elements of information theory*. John Wiley & Sons: 2012.

30. Frenkel, D.; Smit, B., *Understanding molecular simulation: from algorithms to applications*. Elsevier: 2001; Vol. 1.

31. Vinyals, O.; Blundell, C.; Lillicrap, T.; Wierstra, D. In *Matching networks for one shot learning*, Advances in neural information processing systems, 2016; pp 3630-3638.

32. Bartunov, S.; Vetrov, D. P., Fast adaptation in generative models with generative matching networks. *arXiv preprint arXiv:1612.02192* **2016**.

33. Maaten, L. v. d.; Hinton, G., Visualizing data using t-SNE. *Journal of machine learning research* **2008,** *9* (Nov), 2579-2605.

34. Vaswani, A.; Shazeer, N.; Parmar, N.; Uszkoreit, J.; Jones, L.; Gomez, A. N.; Kaiser, Ł.; Polosukhin, I. In *Attention is all you need*, Advances in Neural Information Processing Systems, 2017; pp 5998-6008.

35. Ruder, S., An overview of multi-task learning in deep neural networks. *arXiv preprint arXiv:1706.05098* **2017**.

36. Thrun, S., Lifelong learning algorithms. In *Learning to learn*, Springer: 1998; pp 181-209.

37. Vilalta, R.; Drissi, Y., A perspective view and survey of meta-learning. *Artificial intelligence review* **2002,** *18* (2), 77-95.

38. Bengio, Y.; Louradour, J.; Collobert, R.; Weston, J. In *Curriculum learning*, Proceedings of the 26th annual international conference on machine learning, ACM: 2009; pp 41-48.

39. De Silva, V.; Tenenbaum, J. B. *Sparse multidimensional scaling using landmark points*; Technical report, Stanford University: 2004.

40. Kingma, D. P.; Ba, J., Adam: A method for stochastic optimization. *arXiv preprint arXiv:1412.6980* **2014**.

41. Torrie, G. M.; Valleau, J. P., Nonphysical sampling distributions in Monte Carlo free-energy estimation: Umbrella sampling. *Journal of Computational Physics* **1977,** *23* (2), 187-199.

42. Laio, A.; Parrinello, M., Escaping free-energy minima. *Proceedings of the National Academy of Sciences* **2002,** *99* (20), 12562-12566.

43. Wierstra, D.; Schaul, T.; Glasmachers, T.; Sun, Y.; Peters, J.; Schmidhuber, J., Natural evolution strategies. *The Journal of Machine Learning Research* **2014,** *15* (1), 949-980.

44. Anthony, T.; Tian, Z.; Barber, D. In *Thinking fast and slow with deep learning and tree search*, Advances in Neural Information Processing Systems, 2017; pp 5360-5370.

45. Schütte, C.; Noé, F.; Lu, J.; Sarich, M.; Vanden-Eijnden, E., Markov state models based on milestoning. *The Journal of chemical physics* **2011,** *134* (20), 05B609.

46. Lindorff-Larsen, K.; Piana, S.; Dror, R. O.; Shaw, D. E., How fast-folding proteins fold. *Science* **2011,** *334* (6055), 517-520.

47. Frauenfelder, H.; Sligar, S. G.; Wolynes, P. G., The energy landscapes and motions of proteins. *Science* **1991,** *254* (5038), 1598-1603.

48. Krizhevsky, A.; Sutskever, I.; Hinton, G. E. In *Imagenet classification with deep convolutional neural networks*, Advances in neural information processing systems, 2012; pp 1097-1105.

49. Mikolov, T.; Karafiát, M.; Burget, L.; Černocký, J.; Khudanpur, S. In *Recurrent neural network based language model*, Eleventh annual conference of the international speech communication association, 2010.